\documentstyle[11pt,paspconf,psfig,twoside]{article}
\input{gcbook.pgi}
\begin{document}
%\begin{center}
%{\footnotesize to appear in: "The Central Parsecs of the Galaxy", eds. H. Falcke, A. Cotera, \\W. Duschl, F. Melia, M. Rieke, ASP Conf. Series}
%\end{center}
\title{The Jet Model for Sgr A*}
\author{H. Falcke}
\affil{Max-Planck-Institut f\"ur Radioastronomie, Auf dem H\"ugel 69,
D-53121 Bonn, Germany}
\affil{Steward Observatory, The University of Arizona, Tucson, AZ 85721}

\begin{abstract}
In this paper the jet model for the supermassive black hole candidate
Sgr A* in the Center of the Galaxy is reviewed. The most recent model,
with a reduced set of parameters, is able to account for all major
radio properties of the source: size, structure, flux density, and
spectrum. The model requires a minimum jet power of $\sim10^{39}$
erg/sec and in a symbiotic jet/disk system implies a minimum accretion
rate of a few times $10^{-8} M_\odot$/yr for a radio loud jet or
$\sim10^{-5} M_\odot$/yr for a radio quiet jet.  Low near-infrared
limits on the Sgr A* flux then imply that the accretion flow onto the
central black hole must be radiatively deficient, but most likely has
a high viscosity. Within the jet model the high-frequency part of the
Sgr A* spectrum is self-consistently explained as the nozzle of the
outflow. In a symbiotic model this innermost region of the jet could
possibly be identified with the innermost region of an advection
dominated accretion disk, a Bondi-Hoyle accretion flow, or any other
type of under-luminous accretion process. The compact nozzle region is
of particular importance since it can be used as a background photon
source against which the central black hole could be directly imaged
with future mm-VLBI experiments.
\end{abstract}


\section{Introduction}
Sgr~A* has been at the center of an extensive discussion at this
workshop. An overwhelming number of observations now demonstrate that
this compact radio source in the center of the Galaxy is associated
with a dark mass of $2.6\cdot10^6 M_\odot$ (Haller et al.~1996; Ghez
et al.~1998 \& 1999; Eckart \& Genzel 1999; Zhao et al.~1999). In the
past Sgr A* was only detected in the radio and many properties were
associated with it only because of it being the most unusual radio
object in the entire region (see Falcke 1996a for a review). Now we
have much more direct information about its various properties. For
example Ghez et al.~(1999) show that the central dark mass
concentration does indeed peak at the position of Sgr~A* without any a
priori assumptions about its nature. Stolovy et al.~(1999) find that
the only short-term variable source in their NICMOS near-infrared
images of the very center also coincides with Sgr A*. Eckart \& Genzel
(1999) also claim the detection of a variable NIR source as a
counterpart to Sgr A*.  Cotera et al.~(1999) report a detection of an
unusual and narrow mid-infrared ridge at the position of Sgr A* but it
is unclear how and if that is related to the central dark mass
concentration.

While the amount of dark mass itself and its association with Sgr~A*
now seems well established, the nature of Sgr~A* is not. Any model
needs to explain the compact size, the possibly elongated structure
(Lo et al.~1998), the inverted radio spectrum, the cut-off of the
spectrum towards the infrared, and the faintness of emission at other
wavelengths. At present it is still completely unclear which of the
proposed emission models, if any, would be a fair description of the
actual processes in the Galactic Center. Models proposed so far
include advection dominated two-temperature accretion flows (ADAFs,
Narayan et al.~1995), Bondi-Hoyle accretion (Melia 1994), emission
from mono-energetic electrons (Duschl \& Lesch 1994), or a jet
(Falcke, Mannheim, \& Biermann 1993). Even though they involve
different physical assumptions, all models usually assume the presence
of a supermassive black hole in Sgr A*; a comparison of the various
approaches is given in Falcke (1996a) and Melia (1999). This paper
will now mainly focus on the jet model and its recent improvements.

\section{The Jet Model}
The jet model originally proposed by Falcke et al.~(1993) assumes that
the radio emission from Sgr A* can be explained as emission from a
radio jet, similar to, but with much lower power than, those seen in
other nearby active galaxies or even quasars. Since the original
publication the model has been refined further in Falcke (1996b) and
Falcke \& Biermann (1999). `Refined' in this context means that the
model was further simplified to have fewer free parameters:
equipartition factors were set to unity, the electron distribution was
assumed to be mono-energetic\footnote{In contrast to Duschl (1999) the
mono-energetic distribution here is not an essential ingredient of the
model since the final spectrum will result from a superposition of
many self-absorbed components.}, and the velocity field of the jet was
calculated self-consistently.

The basic idea is to start with a certain power $Q_{\rm jet}$ which is
used to accelerate plasma through a nozzle of characteristic scale
height $Z_{\rm nozz}$. The plasma is treated as an almost relativistic
proton/electron gas with a sound speed of $\simeq0.4c$ and an
adiabatic index $\Gamma=4/3$. The magnetic field is assumed to be
tangled and in equipartition with relativistic particles. Moreover,
the combined energy of relativistic particles and magnetic field is
assumed to be in equipartition with the kinetic energy of the outflow.
With these equipartition assumptions the particle density and magnetic
field energy densities then follow directly from the input parameter
$Q_{\rm jet}$ at the sonic point inside the nozzle.

For the purpose of calculating the radio emission one can treat
the acceleration region as a black box which pushes the plasma through
the sonic point. In a one-dimensional treatment the flow can then be
described as a pressure-driven wind with the jet proper velocity
$\gamma_{\rm j}\beta_{\rm j}$ given by the Euler equation to be

\begin{equation}\label{euler2}\label{v}
\left({\left({\Gamma+\xi\over\Gamma-1}\right)(\gamma_{\rm j}\beta_{\rm
j})^2-\Gamma\over\gamma_{\rm j}\beta_{\rm j}}\right)
{\partial\gamma_{\rm j}\beta_{\rm j}\over\partial z}
={2\over z}
\end{equation}
and with $\xi=\left(\gamma_{\rm j}\beta_{\rm
j}/(\Gamma(\Gamma-1)/(\Gamma+1))\right)^{1-\Gamma}$ (Falcke 1996b).

Assuming a vacuum, the jet will expand transversally with roughly
sound speed after leaving the nozzle. The resulting pressure gradient
along the jet (Fig.~1, left) will accelerate the flow to
bulk Lorentz factors of a few (Fig.~1, right). This is of
course an extremely simplified treatment and if there are additional
acceleration processes (centrifugal, magnetic) the velocity would be
higher still. The resulting shape of the outflow is roughly conical
and depicted in Figure~2 (left). The axial ratio is of
the order 3 to 4:1.

\begin{figure}[h]\label{velocity}\label{pressure}
 \centerline{\psfig{figure=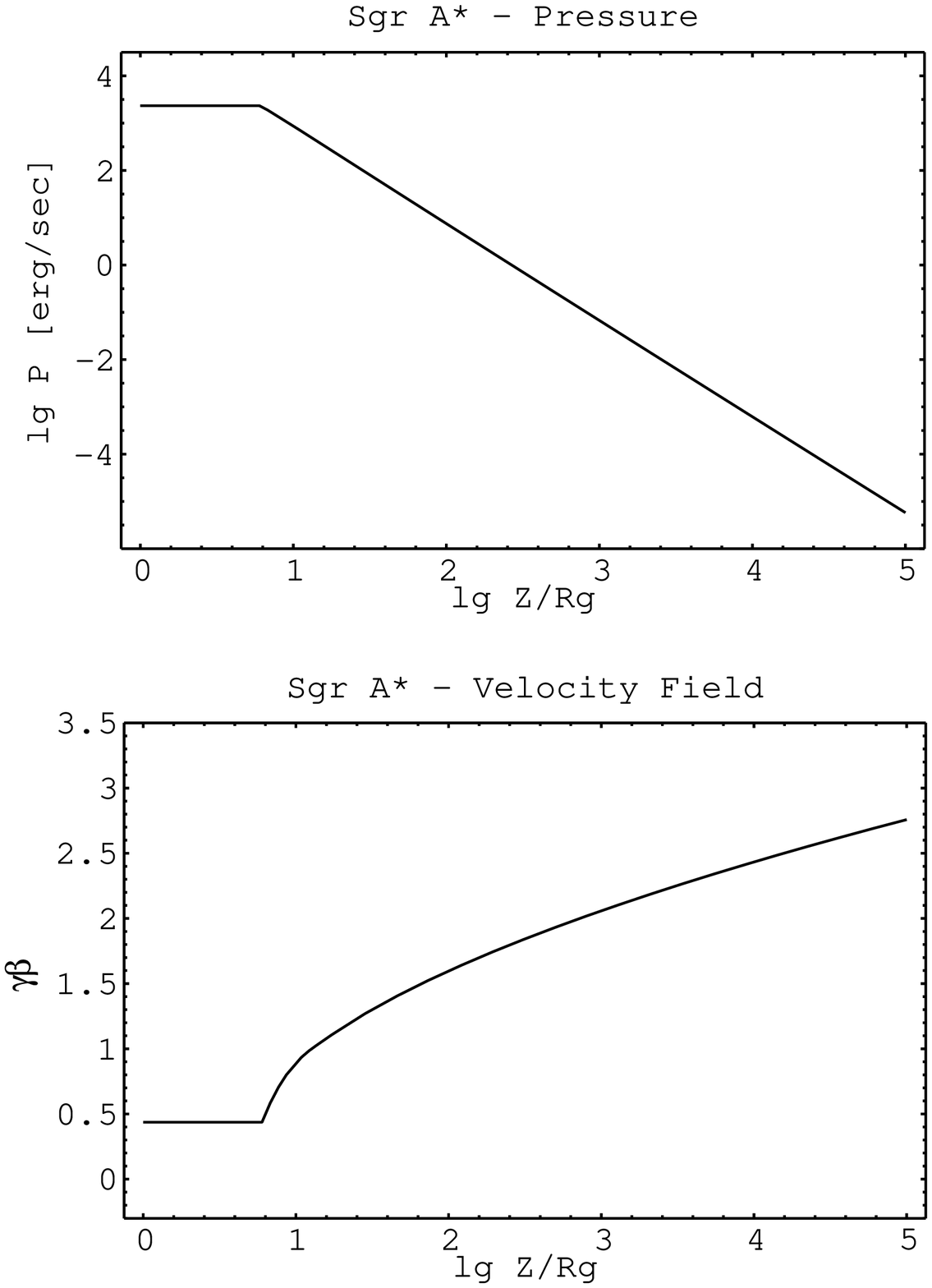,width=0.475\textwidth,bbllx=1.6cm,bblly=15.0cm,bburx=18.1cm,bbury=26.2cm,clip=}\hfill\psfig{figure=sgrvelocity.drw.ps,width=0.49\textwidth,bbllx=1.2cm,bblly=3.0cm,bburx=18.1cm,bbury=14.0cm,clip=}
}
\caption[]{Run of pressure (left) and proper velocity (right) in the
Sgr A* jet as a function of radius (in gravitational radii $GM_\bullet
c^{-2}$) calculated for a simple pressure driven outflow of
relativistic plasma.}
\end{figure}

Based on this velocity field, density distribution, and geometry it is
then possible to calculate the expected radio spectrum of
Sgr~A*. Here, another free parameter is the energy distribution of the
radiating particles (electrons or pairs). The simplest approach is to
use a quasi mono-energetic electron distribution with a characteristic
energy $\gamma_{\rm e}$ (Duschl \& Lesch 1994). However, any other
distribution with a similar characteristic energy (e.g., power-law with
low-energy break or even thermal distributions) would lead to
similar results.

\begin{figure}[h]\label{spectrum}\label{geometry}
\centerline{\psfig{figure=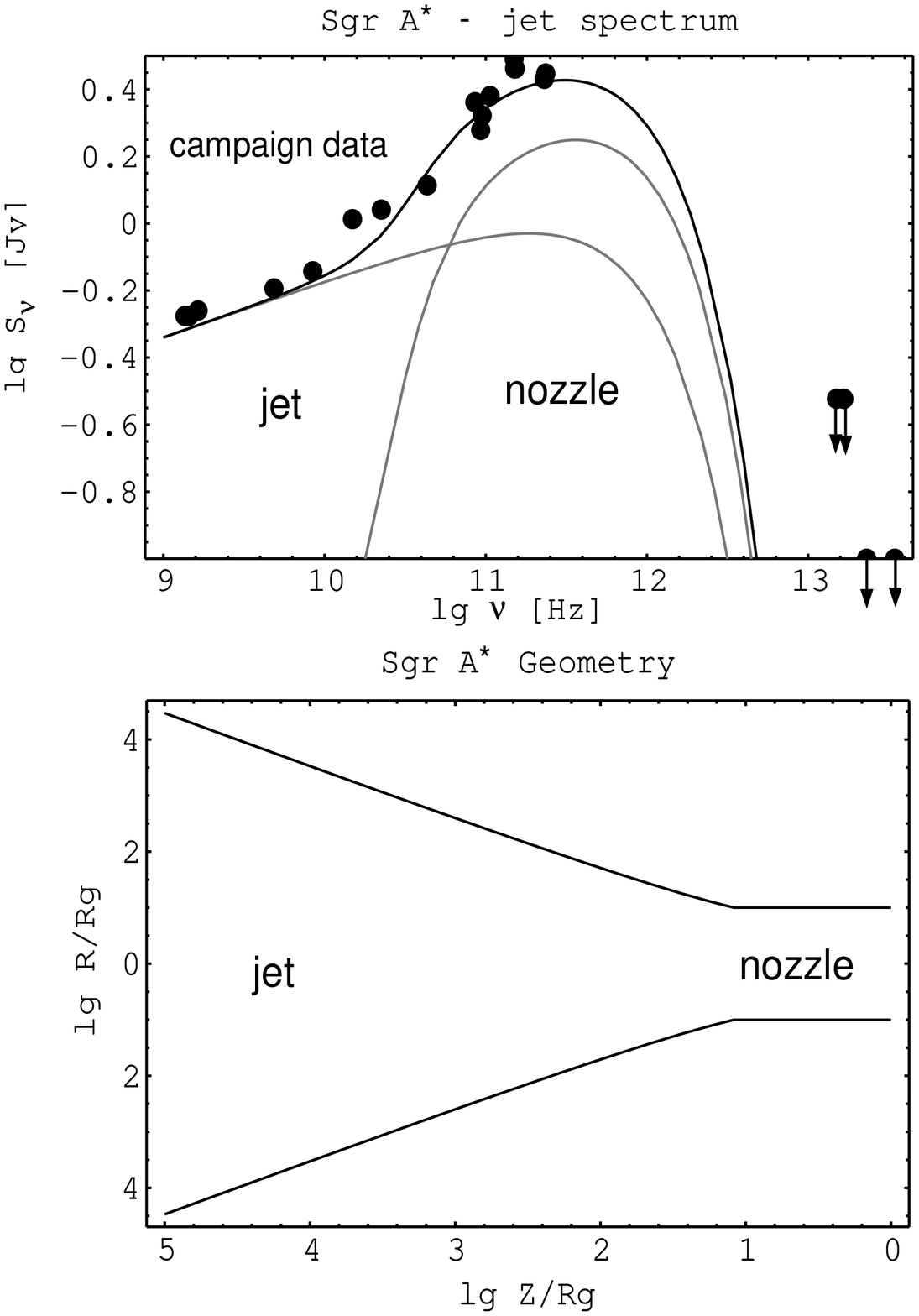,width=0.46\textwidth,bbllx=1.2cm,bblly=1.8cm,bburx=18.2cm,bbury=14.1cm,clip=}%bbllx=2.6cm
\hfill\psfig{figure=sgrspec.drw.ps,width=0.49\textwidth,bbllx=1.3cm,bblly=14.4cm,bburx=18.3cm,bbury=25.9cm,clip=}}
\caption[]{Left: geometry of the jet model for Sgr A*. Note
the logarithmic scale on the vertical axis ending at and not going
through zero. Right: radio spectrum of Sgr A*. The radio data are from
Falcke et al.~(1998) and references therein. The solid line is the
prediction of the jet model, where the gray lines indicate which parts
are contributed by the jet and the nozzle respectively.}
\end{figure}

As shown in Falcke \& Biermann (1999) all these simplifications yield
the following expressions for the observed flux density and angular
size of Sgr A* observed at a frequency $\nu$ as a function of jet
power. For a source at a distance $D$, with black hole mass
$M_\bullet$, size of nozzle region $Z_{\rm nozz}$ (in $R_{\rm g}=G
M_\bullet/c^2$), jet power $Q_{\rm jet}$, inclination angle $i$, and
characteristic electron Lorentz factor $\gamma_{\rm e}$, the observed
flux density spectrum is given as

\begin{eqnarray}\label{flux}
S_{\nu}&&=10^{3.03\cdot\xi_0}\;{\mbox mJy}\;\left({Q_{\rm jet}\over10^{39} \mbox{erg/sec}}\right)^{1.27\cdot\xi_1}
\nonumber\\&&\cdot
\left({D\over10{\rm kpc}}\right)^{-2}
\left({\nu\over8.5 {\rm GHz}}\right)^{0.20\cdot\xi_2}
\left({M_\bullet\over2.5\cdot10^6 M_\odot}{Z_{\rm nozz}\over10 R_{\rm g}}\right)^{0.20\cdot\xi_2}
\nonumber\\&&\cdot
\left(3.9\cdot\xi_3\left(\gamma_{\rm e,0}\over200\right)^{-1.4\cdot\xi_4}
-2.9\cdot\xi_5\left(\gamma_{\rm e,0}\over200\right)^{-1.89\cdot\xi_6}\right),
\end{eqnarray}
with the correction factors $\xi_{0-6}$ depending on the
inclination angle $i$ (in radians):

\begin{eqnarray}
\xi_0&=&2.38 - 1.90\,i + 0.520\,{i^2}\\
\xi_1&=&1.12 - 0.19\,i + 0.067\,{i^2} \\
\xi_2&=&-0.155 + 1.79\,i - 0.634\,{i^2} \\
\xi_3&=&0.33 + 0.60\,i + 0.045\,{i^2} \\
\xi_4&=&0.68 + 0.50\,i - 0.177\,{i^2} \\
\xi_5&=&0.09 + 0.80\,i + 0.103\,{i^2}\\
\xi_6&=&1.19 - 0.29\,i + 0.101\,{i^2}.
\end{eqnarray}
Likewise, the characteristic angular size scale of the emission region
is given by

\begin{eqnarray}\label{size}
\Phi_{\rm jet}&=&4.7\cdot\chi_0\,\mbox{mas}\,\sin{i}
\nonumber\\&\cdot&
\left(\gamma_{\rm e,0}\over200\right)^{1.77\cdot\chi_1}
\left({D\over10{\rm kpc}}\right)^{-1}
\left({\nu\over8.5 {\rm GHz}}\right)^{-0.89\cdot\chi_1}
\nonumber\\&\cdot&
\left({Q_{\rm jet}\over10^{39} \mbox{erg/sec}}\right)^{0.60\cdot\chi_1}
\left({M_\bullet\over2.5\cdot10^6 M_\odot}{Z_{\rm nozz}\over10 R_{\rm g}}\right)^{0.11\cdot\chi_2},
\end{eqnarray}
with the angle dependent correction factors

\begin{eqnarray}
\chi_0&=& 4.01 - 5.65\,i + 3.40\,{i^2} - 0.76\,{i^3}\\
\chi_1&=& 1.16 - 0.34\,i + 0.24\,{i^2} - 0.059\,{i^3}\\
\chi_2&=& -0.238 + 2.63\,i - 1.85\,{i^2} + 0.459\,{i^3},
\end{eqnarray}
where again the inclination angle $i$ is in radians. All correction
factors are normalized to yield unity at $i=1$ rad ($\sim57^\circ$).

For Sgr~A* we are left with four basic unknown quantities in this
equation: the inclination angle $i$, the jet power $Q_{\rm jet}$, the
characteristic electron Lorentz factor $\gamma_{\rm e}$, and the
relative nozzle size $Z_{\rm nozz}$. The expectation from fitting of
other radio cores is that $Z_{\rm nozz}$ will be close to the jet
origin of order 10 $R_{\rm g}$ and $\gamma_{\rm e}$ is of order
$10^2$. Moreover, given our position in the Galaxy it seems unlikely
that Sgr~A* would be pointing right at us and the inclination angle
should probably be larger than say $30^\circ$. 

If one assumes that the mm-bump in the spectrum is in fact intrinsic
to Sgr~A* and is produced by the innermost region of the jet/accretion
disk (i.e., the nozzle) one can use this feature to effectively
constrain $\gamma_{\rm e}$ and $Z_{\rm nozz}$ for a given $Q_{\rm
jet}$. In addition, for a given flux density and size of the core at
frequencies below the bump, one can also solve for the jet power and
the inclination angle.

Unfortunately, the exact size of Sgr A* is still uncertain. Lo et
al.~(1998) claim that the intrinsic size of Sgr A* is $\sim0.45$ mas
at 43 GHz for a flux density of 1.1 Jy. Even though these numbers
should be considered to be very tentative at present (see Bower et
al.~1999a for a discussion of the uncertainties involved), we will use
these number to fit the jet model. Figure~2 shows the predictions of
the jet model for the Sgr~A* radio spectrum in comparison to the
data. The overall spectrum including the submm-bump is well reproduced
for a nozzle size of $Z_{\rm nozz}=12 R_{\rm g}=35 \mu$as (needs to be
multiplied by two if we see both sides), a characteristic electron
Lorentz factor of 125, an inclination angle of $\sim45^\circ$, and a
jet power of $Q_{\rm jet}\ga10^{38.7}$ erg/sec.  The predicted
spectral index of the low-frequency part of the spectrum is
$\alpha\simeq0.17$ ($S_\nu\propto\nu^\alpha$) which agrees well with
the observed one. The predicted size as a function of frequency
roughly scales as $\nu^{-0.9}$ and is shown in Fig.~3 together with
results from recent mm-VLBI experiments. As far as the observations
are concerned, one has to point out that the measurements at
$\lambda$3 and $\lambda$1.3~mm do not probe the NS direction, and hence
for a jet geometry along the axis seen by Lo et al.~(1998) could
underestimate the true size of Sgr A* (see Doleman et al.~1999 for a
discussion of this important detail).

\begin{figure}[ht]\label{size2}
\centerline{\psfig{figure=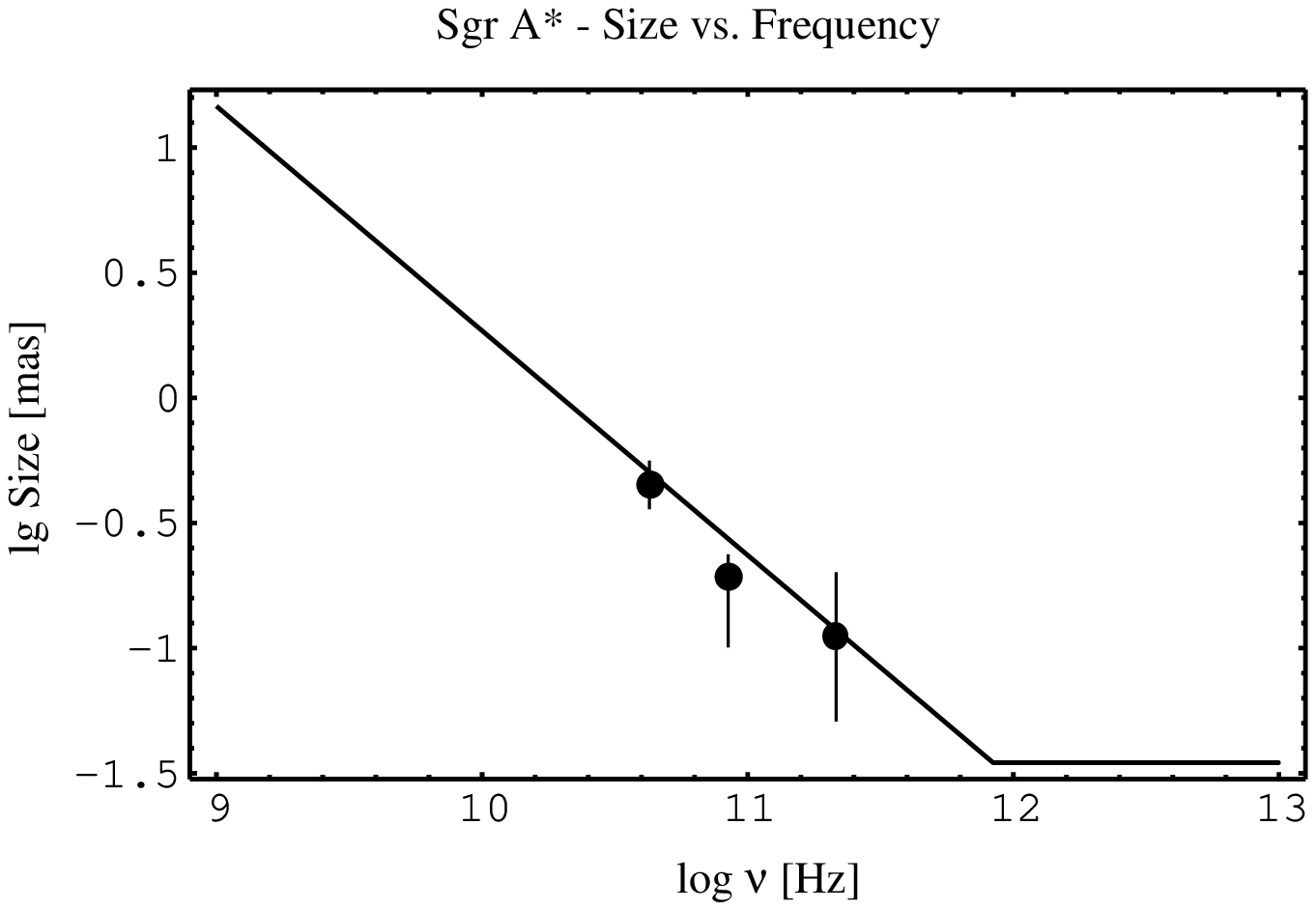,width=0.9\textwidth,bbllx=3.1cm,bblly=16.4cm,bburx=16.8cm,bbury=26.2cm}}
\caption[]{Intrinsic size of Sgr A* (major axis, probably north-south)
versus frequency as predicted in the jet model (solid line). The
parameters ($i, \;\gamma_{\rm e}, \;Q_{\rm jet}$) have been chosen to
match the intrinsic NS size of 0.44 mas at $\lambda$7mm given by Lo et
al.(1998). The data points at $\lambda$3 and $\lambda$1.3 mm were
taken from Krichbaum et al.~(1999) and Doleman et al.~(1999). The
error bars indicate the reported range of values. Note that the sizes
at $\lambda$3 and $\lambda$1.3 mm are not for the NS direction.}
\end{figure}

\section{Discussion}
\subsection{Results of the Jet Model}
In general it can be said that the jet model explains all known
observational details of Sgr A* reasonably well. This shows that the
claim by Lo et al.~(1998) that none of the existing model could
reproduce their data is unfounded. In fact, all the observational
results Lo et al. claim (i.e., elongation, intrinsic size, intrinsic
size scaling with frequency) had already been well predicted earlier
by the jet model within reasonable margins. Nevertheless, one needs to
point out that the core size within the current version of the jet
model cannot be very accurately predicted since it depends quite
sensitively on the electron Lorentz factor, leaving an inherent
uncertainty of a few.

Another important benefit of the jet model is that it not only fits a
single source, Sgr A*, but the same relatively simple equations
presented here can be used to explain in detail well-constrained radio
cores in other galaxies like M81 or NGC4258 and even the galactic
superluminal jet source GRS1915+105 (Falcke \& Biermann 1996, 1999).

\subsection{The Accretion Flow in Sgr A*}
Moreover, the jet model allows one to make interesting inferences for
the nature of the accretion process in Sgr A*. Our current
understanding is that jets are formed in the innermost region of an
accretion disk near a compact object. This seems to be true for young
stars as well as for neutron stars and black holes. Jet and disk
therefore cannot be considered in isolation but form a symbiotic
system (Falcke \& Biermann 1995). In many cases one finds that the jet
power is a significant fraction of the total accretion power and is
comparable to the accretion luminosity.

Sgr~A* adds another twist to this story. Since the jet model used here
basically assumed the most efficient type of radio jet, we can place a
firm lower limit on the jet power. Any modifications to the model,
other than assuming unreasonably small inclination angles, will only
lead to an increased energy demand and a larger $Q_{\rm jet}$. If we
assume that half of the energy released in an accretion process onto a
rotating black hole is used to drive the jet, we need a {\it minimum }
accretion rate of $5\cdot10^{-8}M_\odot$. Interestingly, any standard
accretion disk model with such an accretion rate would violate the
current upper limits on the NIR emission of Sgr~A* as presented for
example in this volume. Hence, we must conclude that the accretion
disk in Sgr A* must be radiatively deficient, e.g., is optically thin
and advection dominated (Narayan et al.~1995). As discussed in Falcke
\& Heinrich (1994) the disk in Sgr~A* would turn optically thin when
the viscosity parameter is $\alpha\ga10^{-3}$. The jet model therefore
also indirectly suggests that a relatively efficient viscosity
mechanism is at work in Sgr~A*.

Unfortunately, it is not easy to give a similarly robust estimate for
an upper limit of the accretion rate. From studies of other AGN (see
discussion in Falcke et al.~1995; Falcke \& Biermann 1995) we know
that there is a dichotomy of the radio-luminosity of jets, the nature of
which is completely unknown. At least in quasars the radio cores of
radio-quiet jets seem to be roughly a factor hundred less luminous for
a given accretion rate than radio-loud jets. Assuming that Sgr A*
would physically resemble such a radio-quiet quasar core, would imply
that in fact the accretion rate needed to power Sgr A* is of the order
$10^{-5} M_{\sun}$/yr. This is close to the value needed in the ADAF
model, though not quite yet the value predicted by 3D hydrodynamic
simulations of the GC region (Melia 1999; Coker \& Melia 1999). In either case
a radio-quiet jet model would still require a very high `inefficiency
factor' for the production of thermal radiation to explain the low NIR
limits for Sgr A*.

The idea of the jet/disk-symbiosis could be pushed even further for
Sgr~A*: an interesting part of the jet model is the nozzle, which is
supposed to be the innermost part of the jet and in more luminous AGN
is usually invisible due to energy-loss arguments. In a symbiotic
system this region would be the interface between the accretion flow
(inflow) and the jet (outflow) and it cannot really be claimed to
belong to either one---the nozzle could be as much part of the disk as
it is part of the jet. This would allow one to easily combine existing
models for the accretion flow (e.g., Melia 1999, Coker \& Melia 1999; Narayan et
al.~1995) with the jet model, by dropping the assumption that these
models explain the cm-spectrum and assuming that their mm-emission
components are in fact the black box labeled 'nozzle' in the jet
picture. An ADAF or Bondi-Hoyle accretion flow with outflow in its
innermost region might do just that and explain what the nozzle really
is.

\subsection{Predictions}
A number of predictions from the jet model can be made that can be
tested in the near future. Sgr~A* should become resolved at 3 and 1 mm
in the NS direction once a suitable mm-VLBI array is available. From
analogy to other radio cores one would expect a polarization at the
percent level at mm-wavelengths where interstellar propagation effects
become negligible (Bower et al.~1999a\&b). The most likely direction
of the magnetic field is probably along the jet axis (NS?). Because
the outflow travels from small to large scales and from small to large
wavelengths one would expect that radio outbursts appear first at high
frequencies and then propagate to longer wavelengths. The time scale
for this delay could be relatively short. The model also predicts a
certain level of x-ray emission, since the relativistic electrons in
the nozzle will inverse-Compton scatter their own synchrotron radiation
into the soft x-ray regime. The luminosity, however, will be
relatively low, of the order $\la10^{34}$ erg/sec with a relatively
unusual, curved spectrum. A more detailed calculation was presented
already in Beckert \& Duschl (1997).

\section{Outlook}

\begin{figure}[ht]\label{bh}
\centerline{\psfig{figure=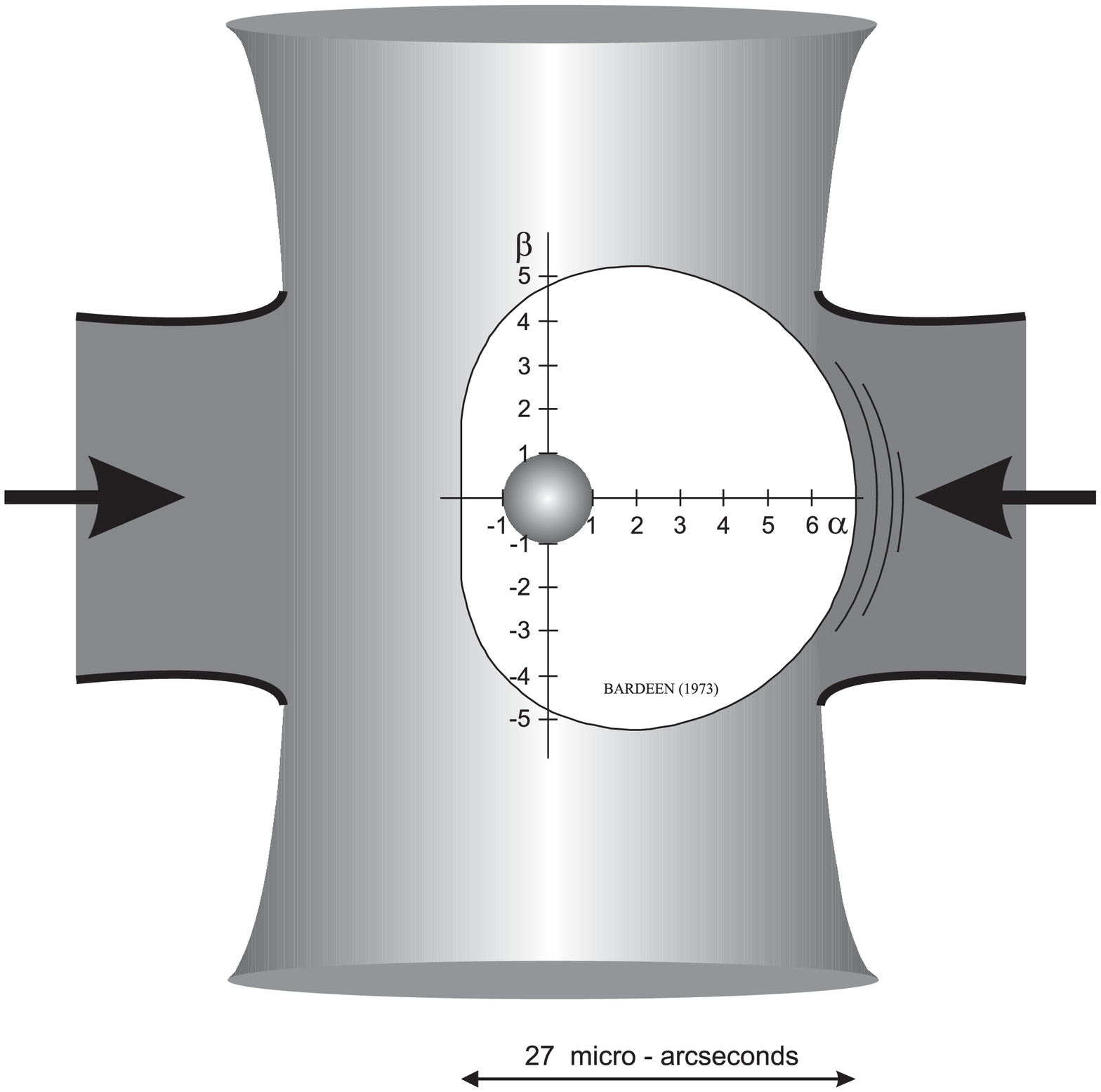,width=0.75\textwidth}}
\caption[]{Sketch of the inner region of Sgr A* with accretion flow
and nozzle surrounding the black hole. Overlaid is an appropriately
scaled reproduction of a Figure from Bardeen (1973), showing the
`hole' of photons absorbed by the black hole if observed against a
background source. A similar process could apply to Sgr A* and its
compact, high-frequency emission component.
}
\end{figure}

The most intriguing consequence of the radio spectrum and its modeling
is the suggestion that the mm-part of the spectrum corresponds to an
ultra-compact region in Sgr~A*. This follows from a few simple
arguments (e.g., Falcke et al.~1998) as well as from basically all
models proposed. Within the jet model, the size of the component used
here to fit the spectrum corresponds to 35$\mu$as. This is
interestingly close to the suspected black hole --- six
Schwarzschild radii. Bardeen (1973) made an interesting calculation,
where he showed that photons from a background source behind a
rotating black hole would be absorbed within an asymmetric disk of 4.5
Schwarzschild radii diameter. Even though the mm-emission region in Sgr
A* is not at infinity behind the black hole one would expect a similar
effect here. Figure 4 shows the Bardeen absorption disk overlayed on a
sketch of Sgr~A* with accretion disk, jet and black hole. Future 220
GHz VLBI experiments---first results of which have already been
reported (Krichbaum et al.~1999)---will have enough resolution to
start imaging such effects directly. At this frequency scattering will
also not be major problem any more. This experiment could provide the
final proof, not only for the validity of a specific model, but also
for the existence of black holes in general.

\acknowledgments This work was supported by the Deutsche
Forschungsgemeinschaft, grants Fa 358/1-1\&2.

\end{document}